\documentclass[aps,twocolumn,a4paper,showkeys,showpacs,footinbib,superscriptaddress]{revtex4-1}
\usepackage{amsmath,amssymb,bm,amsthm}
\usepackage{subfigure}
\usepackage{graphicx}
\usepackage{dcolumn}
\usepackage{bm}
\usepackage[mathlines]{lineno}
\usepackage{booktabs}
\usepackage{color}
\newcounter{one}
\usepackage{url}

\usepackage{textcase}

\usepackage{colortbl}
\usepackage{tabularx}
\usepackage{verbatim}
\usepackage{multirow}

\usepackage[T1]{fontenc} 
\usepackage{lmodern}
\usepackage{bbm}
\usepackage[utf8]{inputenc}
\usepackage{amsfonts}
\usepackage{array}

\usepackage{bbm}
\usepackage{enumitem}
\usepackage{umoline}
\usepackage[usenames,svgnames]{xcolor}
\usepackage{natbib}
\usepackage[hyperindex,breaklinks]{hyperref}
\hypersetup{
     colorlinks=true,       		
     linkcolor=Navy,          	
     citecolor=Navy,            
     filecolor=Navy,      		
     urlcolor=Navy,           	
    runcolor=cyan,
 }
\usepackage{amsmath,amssymb,bm,amsthm}

\def\beq{\begin{equation}}
\def\eeq{\end{equation}}
\def\nbeq{\begin{equation*}}
\def\neeq{\end{equation*}}
\def\beqa{\begin{eqnarray}}%
\def\eeqa{\end{eqnarray}}%

\def\<{\langle}
\def\>{\rangle}

\def\Tr{{\rm Tr}}

\def\flo{{\rm F}}

\newcommand{\sectionprl}[1]{{\par\it #1.---}}
\begin{document}
\title{Heating in integrable time-periodic systems}

\author{Takashi Ishii}
\email{ishii3@iis.u-tokyo.ac.jp}
\affiliation{
Department of Physics, Graduate School of Science,
University of Tokyo, Kashiwa 277-8574, Japan
}
\author{Tomotaka Kuwahara}
\email{tomotaka.kuwahara@riken.jp}
\altaffiliation{Present address: Mathematical Science Team, RIKEN Center for Advanced Intelligence Project (AIP),1-4-1 Nihonbashi, Chuo-ku, Tokyo 103-0027, Japan}
\affiliation{
Advanced Institute for Materials Research, Tohoku University, Sendai 980-8577, Japan
}
\affiliation{
Department of Physics, Graduate School of Science,
University of Tokyo, Kashiwa 277-8574, Japan
}
\author{Takashi Mori}
\email{mori@spin.phys.s.u-tokyo.ac.jp}
\affiliation{
Department of Physics, Graduate School of Science,
University of Tokyo, Bunkyo-ku, Tokyo 113-0033, Japan
}
\author{Naomichi Hatano}
\email{hatano@iis.u-tokyo.ac.jp}
\affiliation{
Institute of Industrial Science,
University of Tokyo, Kashiwa 277-8574, Japan
}

\begin{abstract}
We investigate a heating phenomenon in periodically driven integrable systems that can be mapped to free-fermion models. We find that heating to the high-temperature state, which is a typical scenario in non-integrable systems, can also appear in integrable time-periodic systems; the amount of energy absorption rises drastically near a frequency threshold where the Floquet-Magnus expansion diverges. As the driving period increases, we also observe that the effective temperatures of the generalized Gibbs ensemble for conserved quantities go to infinity. By the use of the scaling analysis, we reveal that in the limit of infinite system size and driving period, the steady state after a long time is equivalent to the infinite-temperature state. We obtain the asymptotic behavior $L^{-1}$ and $T^{-2}$ as to how the steady state approaches the infinite-temperature state as the system size $L$ and the driving period $T$ increase.
\end{abstract}
\maketitle

\sectionprl{Introduction}
Closed quantum many-body systems driven by a time-periodic field have been studied actively in recent years~\cite{
Bukov_review2015,Polkovnikov_review2011,Eisert_review2014,RevModPhys.89.011004,%
Kohler_review2005}. %
Analysis of the steady states after a long time is one of the important questions in characterizing nontrivial quantum phenomena~\cite{PhysRevB.94.184304,PhysRevLett.116.250401,PhysRevB.96.045125,PhysRevA.94.063647,Shirai2015_condition,Lazarides2014a,Lazarides2014b,D'Alessio2014}. 
Periodically driven quantum systems gather attention both experimentally~\cite{Aidelsburger2013,Aidelsburger2015,jotzu2014experimental,matsunaga2014light,Bordia2017} and theoretically~\cite{
Kayanuma-Saito2008,Nag2014,Grossmann1991,Grossmann1992,Prosen-Ilievski2011,BastidasPRA2012,Shirai2014a,Pineda2014,BastidasPRL2012,Sau2012,Gu2011,Grushin2014,Lindner2011,Torres2014,Rudner2013,Goldman2014,Struck2012,Jiang2011,Kundu2013,potter2017infinite,PhysRevB.95.201115,PhysRevLett.118.157201,PhysRevB.95.134508,PhysRevB.94.235419,Roy2015,PhysRevA.95.033631,1367-2630-18-9-095002} because of its potential of realizing novel physical phases, such as topological phases~\cite{PhysRevLett.116.250401,Aidelsburger2013,Aidelsburger2015,jotzu2014experimental,Gu2011,Grushin2014,Lindner2011,Torres2014,Rudner2013,Goldman2014,potter2017infinite,PhysRevB.95.134508,PhysRevB.94.235419,1367-2630-18-9-095002}, by using simple time-dependent Hamiltonians. 
Since there is no conventional energy conservation in such systems, unlike in isolated static systems, the system may absorb energy from the periodic drive and the heating may break down the nontrivial quantum phases. 
Thus, it is an important question whether the heating occurs and to what extent the system absorbs energy from the driving~\cite{strater2016interband,reitterPRL2017,zengPRB2017}. 
Heating can be understood as the energy relaxation towards the maximum entropy state~\cite{D'Alessio2013,Ponte2015,D'Alessio2014,Lazarides2014b,Kim2014}; from this perspective it is also a fundamental issue of statistical physics, namely the thermalization~\cite{mori2017thermalization,D'Alessio_2016,Polkovnikov_review2011}.

Non-integrability is considered to play an essential role in the heating of driven systems. 
Numerical studies~\cite{Lazarides2014b,D'Alessio2014,D'Alessio2013,Russomanno2012,PhysRevB.96.045422,PhysRevB.93.104203} of relatively small non-integrable systems have claimed that when the driving period $T$ is small enough, a system does not heat up but stays at a finite temperature even after a long time, while when $T$ is large enough the system heats up to (nearly) infinite temperature. 
In the latter cases, the unlimited heating is believed to be related to the divergence of the high-frequency expansion~\cite{Bukov_review2015,D'Alessio2014} of the Floquet effective Hamiltonian, or the Floquet-Magnus (FM) expansion~\cite{CPA:CPA3160070404,Blanes_review2009}.
In other words, the  above two regimes with different extents of heating may be bordered by the divergence point of the expansion. 
The convergence radius of the expansion presumably approaches zero in the limit of infinite system size~\cite{KUWAHARA201696}; hence, 
at any non-zero driving periods, macroscopic non-integrable systems are expected to heat up to infinite temperature in the long-time limit, although the time-scale may be extremely long~\cite{PhysRevB.95.014112,PhysRevLett.116.120401,KUWAHARA201696,ho2017bounds,machado2017exponentially,Abanin2017}.

On the other hand, when the total dynamics is integrable,
the system has apparent conserved quantities as many as the degrees of freedom of the system~\footnote{Although there is no generally established definition of integrability in quantum systems~\cite{sutherland2004}, in the present paper we consider systems that can be mapped to the free-fermion system as a specific integrable model. }. 
The quantum dynamics should be restricted in a state space characterized by them. 
This leads to the expectation that the unlimited heating does not occur and the system converges to a nontrivial steady state~\cite{Lazarides2014a,haldar2017dynamical,1742-5468-2016-7-073101,gritsev2017arxiv,PhysRevB.94.214301,PhysRevE.93.062119}. 
Indeed, there has been no report on heating to infinite temperature for integrable time-periodic systems, without additional conditions such as a random noise~\cite{PhysRevX.7.031034}. 
Nonetheless, the connection between the heating and the integrability has been only intuitive, and more elaborate analysis is required.

In the present letter, we report a possible scenario of the unlimited heating in periodically driven systems with integrability.  
Our main conclusion is that heating to infinite temperature can occur even with the integrability, but only in an asymptotic sense. 
We consider a free-fermion system and numerically observe how much energy the initial state absorbs from the driving. 
A qualitative behavior of the heating changes drastically around the driving period where the FM expansion diverges; for shorter periods, the system does not heat up infinitely and remains in finite-temperature states. 
Our observation shows that the steady state approaches the infinite-temperature state \textit{in the limit of the system size and driving period tending to infinity}.
To investigate this point quantitatively, we identify the scaling behavior as to how the steady state approaches to the infinite-temperature state as the system size $L$ and the driving period $T$ increase; as we shall see below, the deviation from infinite temperature decays as $L^{-1}$ and $T^{-2}$.

\sectionprl{Analysis of time-periodic systems}
When the Hamiltonian is periodic in time with the period $T$ as $H(t+T)=H(t),$ the unitary time-evolution operator over a single period
\beq
U_{\mathrm{F}}=\mathcal{T}\exp\left(-i\int^T_0 H(t)dt\right) =:e^{-iH_{\mathrm{F}}T}
\label{eq:UF}
\eeq
defines an effective Hamiltonian $H_{\mathrm{F}}$, which we call the Floquet Hamiltonian.
Here, we denote the time-ordering operator by $\mathcal{T}$.  
At stroboscopic times $t=nT$ with $n$ an integer, the time evolution is described by the static Hamiltonian $H_{\mathrm{F}}$. 
We refer to periodically driven systems with an integrable Floquet Hamiltonian $H_{\mathrm{F}}$ as ``integrable time-periodic system.''

In this letter, we consider  a time-periodic system of the form of the following bilinear fermion Hamiltonian:
\beq
H(t)=\sum_{i,j=1}^L\left(a^{\dag}_i \mathcal{M}_{ij}(t)a_j
+{\rm H.c.}\right),
\label{eq:bilinear}
\eeq
where $a^{\dag}$ and ${a}$ are the creation and annihilation operators, respectively, which satisfy the fermionic commutation relations and $L$ denotes the system size. In this case, the Floquet Hamiltonian $H_{\mathrm{F}}$ is also bilinear and can be mapped to free-fermion systems (see Supplementary~{\uppercase\expandafter{\romannumeral 1}}
), and hence this is an integrable time-periodic system. 
This time-periodic system has $L$ pieces of conserved quantities denoted as
\begin{align}
\hat{\mathcal{I}}_p=f^{\dag}_pf_p
\label{eq:Ip}
\end{align}
for $p=1,2,\dots,L$, where $f^{\dag}_p$ and $f_p$ are eigenmodes of $H_{\rm F}$, namely $H_{\rm F}=\sum_{p=1}^L \epsilon_p f^{\dag}_pf_p$ with $\{\epsilon_p\}_{p=1}^L$ the quasi-energies of $H_{\rm F}$. 
The dynamics is constrained in the Hilbert space which conserves all of $\{\hat{\mathcal{I}}_p\}_{p=1}^L$. 
We have $\sum_{p=1}^Lf_p^{\dag}f_p|\psi(t)\>=N|\psi(t)\>$ for all $t$, where $N$ is the number of modes occupied in the initial state. 
Throughout the paper, we will refer to $N$ as the particle number. 
We define the infinite-temperature state as the uniform mixing of all the states with a fixed particle number $N$. 
That is, the infinite-temperature state, which we denote by ${\bf 1}_{(N,L)}$, is proportional to the projection operator $P_{N,L}$ to the Hilbert space with the particle number $N$.

The amount of the energy absorption is deeply related to the FM expansion, which is an expansion of the Floquet Hamiltonian by the power series of the period $T$~\cite{Blanes_review2009,Bukov_review2015}:
\beq
H_\mathrm{F}(T)=\sum_{n=0}^{\infty} T^n\Omega_n(T).
\label{eq:FM-exp}
\eeq
Each of the terms $\Omega_n(T)$ includes high-order nested commutators of the Hamiltonian $H(t)$ (See Refs.~\cite{bialynicki1969explicit,KUWAHARA201696} for the explicit forms).  
In particular, the first term is equal to the time average over one period of the time-dependent Hamiltonian, which we refer to as $H_{\rm ave}$: 
\begin{align} 
\Omega_0(T)=H_{\rm ave}:= \frac{1}{T}\int^T_0 H(t)dt.
\label{FM-01-1}
\end{align}
Throughout the paper, we refer to the expectation value of the operator $H_{\rm ave}$ as the energy of the periodically driven system.
If the period $T$ is sufficiently small, the Floquet Hamiltonian may be approximated by the average Hamiltonian $H_{\mathrm{F}}\approx H_{\rm ave}$, and therefore the system should remain in a low-energy state if we start from the ground state of $H_{\rm ave}$. 
However, the convergence of the FM expansion~\eqref{eq:FM-exp} is generally not assured for large periods. 
A general sufficient condition for the convergence is given by $\int_0^T H(t) dt \le \pi$~\cite{pechukas1966exponential,karasev1976infinite,blanes1998magnus,moan2008convergence,Blanes_review2009}.
It implies that the convergence is ensured only for $T\lesssim 1/\|H(t)\|$ with $\|\cdots\|$ the operator norm.

When the FM expansion diverges, the Floquet Hamiltonian is no longer close to $H_{\rm ave}$ and higher-order terms become dominant. 
In the periodically driven non-integrable systems, the spectral structure is known to resemble that of a random matrix~\cite{D'Alessio2014,Kim2014}.
This implies that the steady state is given by a random state in the total Hilbert space, namely the infinite-temperature state~\cite{Popescu2006}.

On the other hand,  in the integrable cases, the scenario appears to be rather different. 
The convergence of the FM expansion is ensured for a wider region of $T$ in the integrable cases than in non-integrable systems. 
In the integrable time-periodic system that we consider in this letter, the convergence of the FM expansion is ensured for $T\lesssim 1/\|{\sf M}(t)\|$, where ${\sf M}(t)$ is the $L\times L$ matrix which expresses the single-particle Hamiltonian with its norm $\|{\sf M}(t)\|$ remaining finite even in the thermodynamic limit; see supplementary~{\uppercase\expandafter{\romannumeral 1}}
.
In the integrable cases, the Floquet Hamiltonian $H_\flo$ always commutes with the conserved quantities %
and is far from the random matrix in the total Hilbert space.

The steady state of the system given by Eq.~\eqref{eq:bilinear} is known to be given by the generalized Gibbs ensemble (GGE) dependent on the initial state under some moderate assumptions~\cite{Lazarides2014a}. The ensemble reads
\begin{align}
\hat{\rho}_{\rm GGE}=\mathcal{Z}^{-1}\mathrm{exp}\left(-\sum_{p=1}^L\Lambda_p\hat{\mathcal{I}}_p\right)\label{eq:pge}
\end{align}
with $\mathcal{Z}$ the normalization constant. 
We refer to the coefficients $\{\Lambda_p\}_{p=1}^L$  as ``the effective temperatures'' for the conserved quantities $\{\hat{\mathcal{I}}_p\}_{p=1}^L$. 
The effective temperature $\Lambda_p$  is calculated by equating the expectation values of the conserved quantity $\hat{\mathcal{I}}_p=f^{\dag}_pf_p$ for the initial pure state and the GGE given by Eq.~\eqref{eq:pge} as in
\beq
\<\psi_0|f^{\dag}_pf_p|\psi_0\>=\frac{1}{e^{\Lambda_p}+1},\quad p=1,2, \dots, L,
\label{eq:num_Lam}
\eeq
where $\<\psi(t)|f^{\dag}_pf_p|\psi(t)\>=\<\psi_0|f^{\dag}_pf_p|\psi_0\>$ holds, with $|\psi_0\>$ and $|\psi(t)\>$ denoting the initial state and the state at time $t$, respectively. 

We stress that the steady state of the present system is expected to be given by the GGE as in Eq.~\eqref{eq:pge} for all driving periods regardless of the convergence of the FM expansion, although the specific values of $\{\Lambda_p\}_{p=1}^L$ and forms of $\{\hat{\mathcal{I}}_p\}_{p=1}^L$ should be different for different driving periods. 
Still, the expectation values of physical quantities including the energy $H_{\rm ave}$ after a long time may look like those at infinite temperature when the FM expansion diverges. 
This is what we examine in the present letter.

\begin{figure*}
\centering
\subfigure[%
]{
\includegraphics[clip, scale=0.8]{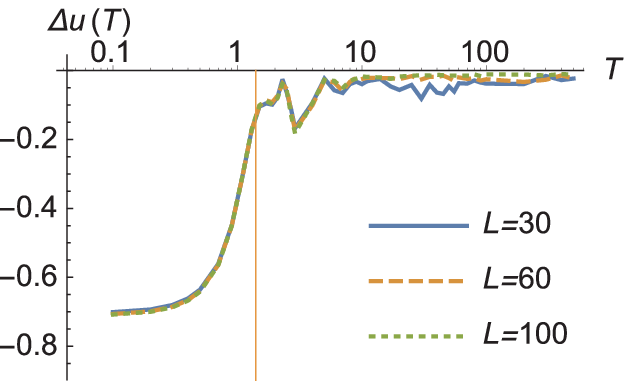}
}
\subfigure[%
]{
\includegraphics[clip, scale=0.8]{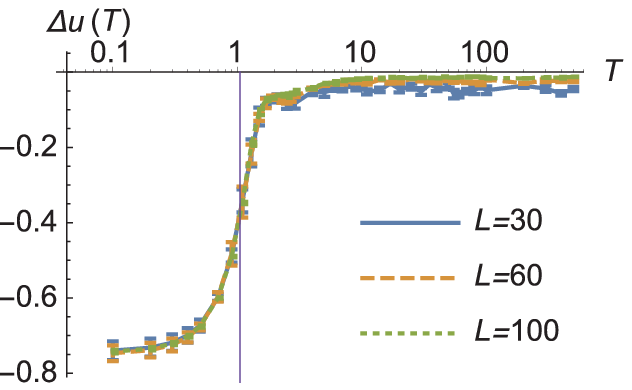}
}
\subfigure[%
]{
\includegraphics[clip, scale=0.8]{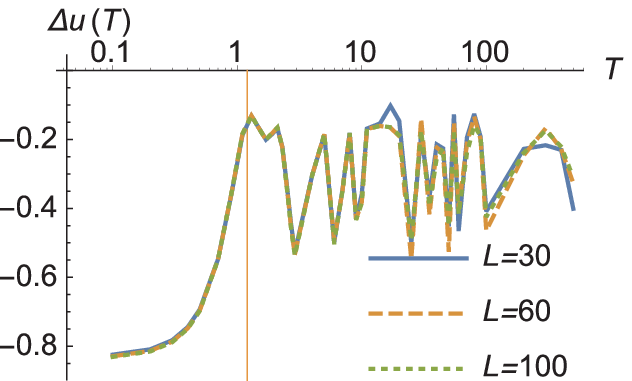}
}\label{Enlstg}
\caption{(color online) Deviation of the energy density from the value at infinite temperature. 
We set $\{h_i\}$ in Eq.~\eqref{XX_random} as (a) the quasi-periodic field, (b) the random Gaussian field with four random samples, and (c) the staggered field; see the description below Eq.~\eqref{XX_random}. 
Each line (and color) shows the result for the corresponding system size. 
The amount of the energy absorption drastically changes around $T\approx 1.$ The vertical line indicates the period where we detected the divergence of the FM expansion.} \label{Fig:Enl}
\end{figure*}

\sectionprl{Model and Setup}
We consider a spin-$1/2$ chain with $L$ sites under the open boundary conditions. 
(We presume that the boundary condition is not critical to the conclusion of our calculation.) 
We periodically switch the system Hamiltonian back and forth between two Hamiltonians $H_1$ and $H_2$. The time evolution operator over one period is
\beq
U_{\mathrm{F}}(T)=e^{-iH_2T/2}e^{-iH_1T/2}.
\label{eq:o_p_u_op}
\eeq
Here, we choose $H_1$ as the $XX$ model Hamiltonian and $H_2$ as the Hamiltonian of the external field along $z$-axis:
\begin{eqnarray}
&&H_1=\sum_{i=1}^{L-1}\left(\sigma_i^x\sigma_{i+1}^x+\sigma_i^y\sigma_{i+1}^y\right),\\
&&H_2=\sum_{i=1}^{L}h_i\sigma_i^z,
 \label{XX_random}
\end{eqnarray}
where we consider three types of $\{h_i\}_{i=1}^L$: a quasi-periodic field $h_i=\sin{\left(2\sqrt{2}\pi\cdot i\right)}$, a random field with $\{h_i\}_{p=1}^L$ given by a random Gaussian with the unit standard deviation, and the staggered field $h_i=(-1)^i$. 
The average Hamiltonian in Eq.~\eqref{FM-01-1} is now given by
\beq
H_{\rm ave}=(H_1+H_2)/2.
\label{aH}
\eeq
After the Jordan-Wigner transformation, both the two Hamiltonians $H_{\rm 1}$ and $H_{\rm 2}$ can be written in bilinear forms of fermionic operators as 
\beq
\sum_{i,j=1}^La^{\dag}_i \mathcal{M}_{ij}a_j.
\label{eq:pn_conserve_bilinear}
\eeq
Therefore, the unitary operator~\eqref{eq:o_p_u_op} defines an integrable Floquet Hamiltonian; see Supplementary~{\uppercase\expandafter{\romannumeral 1}}
.

In order to observe the heating behavior clearly, we choose the ground state of $H_{\rm ave}$ as the initial state $|\psi_0\>$. 
We have numerically confirmed that the particle number $N$ of the initial state is about $L/2$ in the present models. 
We denote the infinite-time average of operators by $\overline{\<\cdots\>}$.
In the following, we consider the infinite-time average of the energy density $\overline{\<H_{\rm ave}\>}/L$ and the effective temperatures $\{\Lambda_p\}_{p=1}^L$ for conserved quantities $\{\hat{\mathcal{I}}_p\}_{p=1}^L$ in \eqref{eq:Ip}.

\sectionprl{Energy density of the steady state}
First we calculate the expectation value of the energy density $H_{\rm ave}/L$ for the steady state after infinite time. 
The infinite-time average $\overline{\<H_{\rm ave}\>}/L$ is given by dropping the off-diagonal terms of $H_{\rm ave}$ represented in the basis of the eigenstates of $H_{\rm F}$~\cite{reimann2008}. We compare $\overline{\<H_{\rm ave}\>}/L$ with the expectation value in the infinite-temperature state ${\bf 1}_{(N,L)}$.

We show in Fig.~\ref{Fig:Enl} the energy density difference
\beq
\Delta u(T):=\overline{\langle H_{\rm ave}\rangle}/L-{\rm Tr}[{\bf 1}_{(N,L)}H_{\rm ave}]/L
\label{edd}
\eeq
against the driving period $T$ for the system sizes $L=30,60,100$. 
The vertical line indicates the period where we numerically detected the divergence of the FM expansion for $H_{\rm F}(T)$; 
see Supplementary Fig.~S1 
 for details. In each panel of Fig.~\ref{Fig:Enl}, we can see a sharp rise of $\Delta u(T)$ around $T\approx 1$, which is close to the divergence point.
However, the size dependence suggests that the energy absorption remains finite above the divergence point $T\gtrsim 1$ in the thermodynamic limit $L\to \infty$.

As the driving period increases, a qualitative difference appears between Figs.~\ref{Fig:Enl}(a,b) and Fig.~\ref{Fig:Enl}(c). 
In the cases of the quasi-periodic and random fields, Figs.~\ref{Fig:Enl}(a) and (b) indicate that the deviation $|\Delta u|$ decays as $T$ and $L$ increase. 
On the other hand, in the case of the staggered field (Fig.~\ref{Fig:Enl}(c)), we clearly see that the infinite-time average of the energy deviates from the infinite-temperature value for all data points.

\begin{figure}
\centering
\subfigure[%
]{
\includegraphics[clip, scale=1]{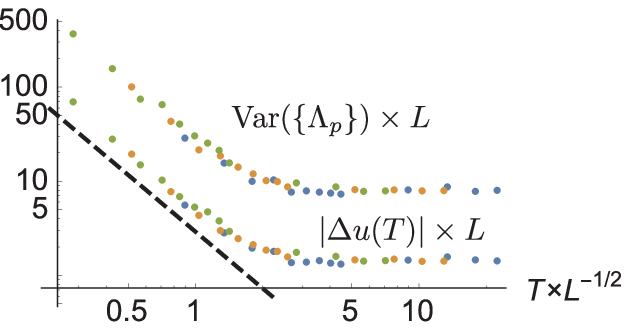}
}
\subfigure[%
]{
\includegraphics[clip,scale=1]{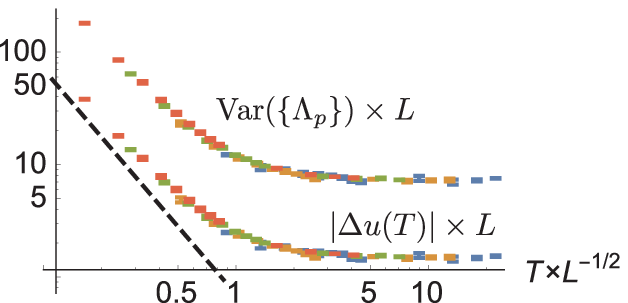}
}
\caption{(color online) Scaling plots of $|\Delta u(T)|$ and ${\rm Var}(\{\Lambda_p\})$. 
We set $\{h_i\}$ in Eq.~\eqref{XX_random} as (a) the quasi-periodic fields and (b) the random Gaussian field with four random samples. 
The driving periods are $T=20,\;30,\;40,\;50,\;60,\;70,\;80,\;90,\;100,\;200,\;300,\;400,\;500$. The system sizes are $L=500,\;1500,\;5000$ for the quasi-periodic field and $L=500,\;1500,\;5000,\;15000$ for the random field. (Different colors indicate different system sizes.) 
The broken line indicates the behavior $T^{-2}.$ }\label{Fig:LaEtoge}
\end{figure}

For the former cases, we calculated $\Delta u(T)$ for larger sizes and wider range of periods than in Figs.~\ref{Fig:Enl} (a,b). 
We found good scaling as in Fig.~\ref{Fig:LaEtoge} (see also Supplementary Fig.~S2
, which shows the data plotted with pre-scaled axes. 
We obtained Fig.~\ref{Fig:LaEtoge} by collapsing the data in Fig.~S2
); the data points lie on a single curve for all $L$ for the shown region of periods when we plot $|\Delta u|\times L$ against $T/\sqrt{L}$. 
(We take the absolute value $|\Delta u|$ instead of $\Delta u$ to plot in the logarithmic scale. 
The scaling breaks in the region with smaller values of $T$.)

This scaling plot means that the quantity $Q:=|\Delta u|$ is given by a scaling function $\tilde Q$ in the form 
\beq
Q(T,L)=L^{-1}\tilde Q(TL^{-1/2}). 
\label{Eq:sc}
\eeq
The curve implies that for a finite system the heating saturates before reaching the infinite-temperature value even in the limit $T\rightarrow\infty$; the part of the curve which is nearly parallel to the horizontal axis corresponds to the region where the saturation occurs~\footnote{For similar scaling analysis, see for example A. -L. Barab\'{a}si and H. E. Stanley, {\it Fractal Concepts in Surface Growth} (Cambridge University Press, 1995) Ch.2}. 
We thereby conclude that the finite-size data should converge to the infinite-size limit $Q=0$ for $T=\infty$ as $Q\propto1/L.$ 
For finite but large $T$, we have $Q(T,L)=T^{-2}[(TL^{-1/2})^2\tilde Q(TL^{-1/2})],$ and hence conclude that $Q\propto T^{-2}$ holds in the infinite-size limit $L\rightarrow \infty.$ 
(The exponent $-2$ corresponds to the gradient of the part of the curve which is not parallel to the horizontal axis. See also Supplementary Fig.~S2
.)
Therefore the system heats up to infinite temperature in the limit $L\rightarrow \infty$ and $T\rightarrow \infty$.

\sectionprl{Effective temperatures for conserved quantities in the GGE}
Next we examine different quantities to confirm that the steady state resembles the infinite-temperature state for observables other than the energy. 
For the purpose, we consider the effective temperatures for the $L$ pieces of conserved quantities $\hat{\mathcal{I}}_p=f^{\dag}_pf_p$, namely $\Lambda_p$ for $p=1,\dots ,L$ given in Eq.~(\ref{eq:pge}). 
For the infinite-temperature state ${\bf 1}_{(N,L)}$, all the expectations $\Tr ({\bf 1}_{(N,L)} \hat{\mathcal{I}}_p )$ for $p=1,2,\ldots,L$ have the same value.
Hence, if all of $\{\Lambda_p\}_{p=1}^L$ in the GGE have the same value, the state \eqref{eq:pge}  reduces to the infinite-temperature state for a fixed particle number $N$.
We therefore analyze the variance among $\{\Lambda_p\}_{p=1}^L$ from the expectation, which we denote as ${\rm Var}(\{\Lambda_p\}):=\frac{1}{L}\sum_{p=1}^L(\Lambda_p-\overline{\Lambda})^2$ with $\overline{\Lambda}:=\frac{1}{L}\sum_{p=1}^L\Lambda_p$. 
The decrease of ${\rm Var}(\{\Lambda_p\})$ means the approach of the steady state to infinite temperature. 

After fixing $L$ and $T$, we can obtain the values of $\{\Lambda_p\}_{p=1}^L$ by numerically computing the left-hand side of Eq.~\eqref{eq:num_Lam}. 
The value $\<\psi_0|f^{\dag}_pf_p|\psi_0\>$ can be computed by expanding $f^{\dag}_p$ and $f_p$ in terms of the eigenmodes of $H_{\rm ave}$.

We conduct the scaling analysis again in order to analyze how ${\rm Var}(\{\Lambda_p\})$ approaches zero as $L$ and $T$ increase. 
As in Fig.~\ref{Fig:LaEtoge}, we find that ${\rm Var}(\{\Lambda_p\})$ follows the same scaling as $|\Delta u|$~\footnote{
In the supplementary Fig.~S3
, we show the $T$-dependence of ${\rm Var}(\{\Lambda_p\})$ for various system sizes $L$.}. 
This reveals that the GGE in Eq.~\eqref{eq:pge} converges to the infinite-temperature state in the limits of $L\rightarrow \infty $ and $T\rightarrow \infty$. 
This gives another piece of evidence for the heating to the infinite-temperature state.

\sectionprl{Discussion}
Here we discuss why a qualitative difference appeared between the cases in which $\{h_i\}$ in $H_2$ was (a) quasi-periodic or (b) random fields, and (c) a staggered field. 
The extent of heating may be explained by the degree of mixing in the one-body state space under the basis of the wave number. 
The first Hamiltonian $H_1$, namely the {\it XX} model, does not mix the state among modes of different wave numbers. 
In order to bring the initial state close to the infinite temperature, it requires for the second Hamiltonian $H_2$ to mix enough the state among modes of different wave numbers. 
In the cases (a) and (b), $H_2$ causes mixing among modes with many different wave numbers, while in the case of (c), $H_2$ only moves the occupation of a mode to another mode with wave number difference $\pi.$
We infer that this is the reason why in the case of (c), no extensive heating occurred as anticipated conventionally, while in the cases of (a) and (b), extensive heating, especially heating to the infinite temperature in the asymptotic sense, occurred in spite of the system being integrable. 
We expect that extensive heating occurs for other driving fields too if the field enough mixes the state among modes of different wave numbers.

Let us comment on the driving-period dependence of the extent of heating in the cases (a) and (b). 
When the driving period $T$ is small enough, $H_{\rm ave}$ approximates the Floquet Hamiltonian, and thus its degree of mixing among modes with different wave numbers is small, resulting in a small amount of heating. 
On the other hand, for the large driving periods where the FM expansion diverges, it should be a challenging problem to evaluate the extent of heating analytically since the Floquet Hamiltonian is determined through the nontrivial contributions of $H_1$ and $H_2$. 
We stress that it is a novel finding in time-periodic systems that scaling behavior in quantum states far from equilibrium exists, having the driving period $T$ as one of its scaling variables.

\sectionprl{Relevance to experiments}
Time-periodic modulation of fermions with nearest-neighbor hopping and quasi-random potential has been achieved in an experiment~\cite{Bordia2017}. 
An alternative way of realizing the present system is to use hard-core bosons, whose Hamiltonian can be mapped to the Hamiltonian of free fermions~\cite{Lazarides2014a}.
Hard-core bosons are also achieved in an experiment of ultracold atoms in optical lattices~\cite{Paredes2004}. 
We thus expect that the scaling of heating expressed by Eq.~\eqref{Eq:sc} should be verified experimentally in these systems.

\sectionprl{Future perspective}
Analytic derivation of the nontrivial scaling exponents in $Q\propto L^{-1}$ and $Q\propto T^{-2}$ remains an open question. 
Our results indicate that the Floquet Hamiltonian $H_{\mathrm{F}}$ resembles a kind of random matrix in the limit of $T\to \infty$ and $L\to \infty$ for the quasi-periodic and random fields. 
The present scalings may be explained by analyzing the difference between the random matrix and $U_{\mathrm{F}}$ for finite $T$ and $L$ if an appropriate quantitative index is found. 
We expect that the scalings demonstrate a universality class and appear in other integrable time-periodic systems. 
Finally, we note that a similar scaling analysis may be helpful also for non-integrable time-periodic systems for further studying its finite-size behavior.

\begin{acknowledgments}
T.I. was partially supported by the Program for Leading Graduate Schools, MEXT, Japan. 
T.K. was partially supported by the Program for World Premier International Research Center Initiative (WPI), MEXT, Japan. 
T.M.'s research was financially supported by JSPS KAKENHI Grant No. 15K17718. 
N.H.'s research was partially supported by Kakenhi Grants Nos. 15K05200, 15K05207, and 26400409 from Japan Society for the Promotion of Science. 
\end{acknowledgments}

\bibliography{itp.bib}

\clearpage
\onecolumngrid
\begin{center}
{\large \bf Supplementary Material for  \protect \\ 
``Heating in integrable time-periodic systems'' }\\
\vspace*{0.3cm}
Takashi Ishii$^{1}$, Tomotaka Kuwahara$^{2,1}$, Takashi Mori$^{3}$ and Naomichi Hatano $^{4}$ \\
\vspace*{0.1cm}
$^{1}${\small \em Department of Physics, Graduate School of Science, University of Tokyo, Kashiwa 277-8574, Japan} \\
$^{2}${\small \em WPI, Advanced Institute for Materials Research, Tohoku University, Sendai 980-8577, Japan} \\
$^{3}${\small \em Department of Physics, Graduate School of Science, University of Tokyo, Bunkyo-ku, Tokyo 113-0033, Japan} \\
$^{4}${\small \em Institute of Industrial science, University of Tokyo, Kashiwa 277-8574, Japan} \\
\end{center}

\setcounter{equation}{0}
\setcounter{figure}{0}
\setcounter{table}{0}
\setcounter{page}{1}
\makeatletter
\renewcommand{\theequation}{S\arabic{equation}}
\renewcommand{\thefigure}{S\arabic{figure}}
\renewcommand{\bibnumfmt}[1]{[S#1]}

\section{One-period unitary operator of time-periodic free fermion systems}\label{Suoff}
Here we show that the calculation of the one-period unitary operator $U_{\rm F}$ of a free-fermion system can be reduced to solving the Floquet problem of an $L\times L$ operator. 
This calculation is necessary for finding the eigenmodes of the Floquet Hamiltonian. 

Consider the time-dependent Hamiltonian 
\beq
H(t)=\sum_{n,m=1}^LM_{nm}(t)a^{\dag}_na_m,
\label{}
\eeq
where $a_n$ is the annihilation operator of a fermion. Denoting the column vector of the annihilation operator as ${\bm a}$, we can express this Hamiltonian as 
\beq
H(t)={\bm a}^{\dag}{\sf M}(t){\bm a},
\label{}
\eeq
where ${\sf M}(t)$ denotes a matrix whose elements are given by $({\sf M}(t))_{nm}=M_{nm}(t)$. 

The one-period unitary operator $U_{\rm F}$ is defined by 
\beq
U_{\rm F}=\mathcal{T}e^{-i\int_0^TH(t)dt}=e^{-iH_{\rm F}T},
\label{}
\eeq
where $\mathcal{T}$ denotes the time-ordering operator. 
The Floquet Hamiltonian is also bilinear as in 
\beq
H_{\rm F} ={\bm a}^{\dag}{\sf M}_{\rm F}{\bm a}.
\label{}
\eeq
In the following we show how to obtain ${\sf M}_{\rm F}$.

We define 
\beq
{\bm a}^{\dag}(t)\equiv U_t{\bm a}^{\dag}U^{\dag}_t,
\label{}
\eeq
where $U_t=\mathcal{T}e^{-i\int_0^tH(s)ds}$. 
Defining ${\sf A}(t)$ as ${\bm a}^{\dag}(t)=:{\bm a}^{\dag}{\sf A}(t),$ we have
\beq
{\bm a}^{\dag}\frac{d{\sf A}(t)}{dt}=\frac{d}{dt}(U_t{\bm a}^{\dag}U_t^{\dag})=-i[H(t),{\bm a}^{\dag}{\sf A}(t)]=-i{\bm a}^{\dag}{\sf M}(t){\sf A}(t)
\label{}
\eeq
with the usage of the equality $[H(t),{\bm a}^{\dag}]={\bm a}^{\dag}{\sf M}(t)$. Therefore we obtain 
\beq
\frac{d{\sf A}(t)}{dt}=-i{\sf M}(t){\sf A}(t).\label{tdA}
\eeq
The solution of Eq.~\eqref{tdA} under the condition $A(0)=1$ is 
\beq
{\sf A}(t)=\mathcal{T}e^{-i\int_0^t{\sf M}(s)ds}.
\label{}
\eeq
Therefore
\beq
{\bm a}^{\dag}(t)=U_t{\bm a}^{\dag}U^{\dag}_t={\bm a}^{\dag}\mathcal{T}e^{-i\int_0^t{\sf M}(s)ds}
\label{}
\eeq
holds.

Now we show how ${\sf M}_{\rm F}$ can be calculated from ${\sf M}(t)$. The basis state of an $N$-particle state is given by 
\beq
a^{\dag}_{i_1}a^{\dag}_{i_2}\dots a^{\dag}_{i_N}|0\>,
\label{}
\eeq
where $i_1, i_2,\dots i_N$ is a set of integers which satisfies $1\leq i_1< i_2<\cdots <i_N\leq L$, and $|0\>$ is the vacuum. We can determine ${\sf M}_{\rm F}$ by observing how the above state is transformed by $U_{\rm F}$. It is expressed as
\beq
U_{\rm F}a^{\dag}_{i_1}a^{\dag}_{i_2}\dots a^{\dag}_{i_N}|0\>=U_{\rm F}a^{\dag}_{i_1}U_{\rm F}^{\dag}\cdot U_{\rm F}a^{\dag}_{i_2}U_{\rm F}^{\dag}\dots U_{\rm F}a^{\dag}_{i_N}U_{\rm F}^{\dag}\cdot U_{\rm F}|0\>.
\label{}
\eeq
 From $U_{\rm F}|0\>=|0\>$ and $U_{\rm F}{\bm a}^{\dag}U_{\rm F}^{\dag}={\bm a}^{\dag}(T)={\bm a}^{\dag}\mathcal{T}e^{-i\int_0^T{\sf M}(s)ds}$, we obtain 
 \beq
U_{\rm F}a^{\dag}_{i_1}a^{\dag}_{i_2}\dots a^{\dag}_{i_N}|0\>=({\bm a}^{\dag}\mathcal{T}e^{-i\int_0^T{\sf M}(s)ds})_{i_1}\cdot ({\bm a}^{\dag}\mathcal{T}e^{-i\int_0^T{\sf M}(s)ds})_{i_2}\cdots ({\bm a}^{\dag}\mathcal{T}e^{-i\int_0^T{\sf M}(s)ds})_{i_N}|0\>.
\label{utrprod}
\eeq
 On the other hand, putting $U_{\rm F}=e^{-i{\bm a}^{\dag}{\sf M}_{\rm F}{\bm a}T}$, we obtain 
 \beq
e^{-i{\bm a}^{\dag}{\sf M}_{\rm F}{\bm a}T}a^{\dag}_{i_1}a^{\dag}_{i_2}\dots a^{\dag}_{i_N}|0\>=({\bm a}^{\dag}e^{-i{\sf M}_{\rm F}T})_{i_1}\cdot ({\bm a}^{\dag}e^{-i{\sf M}_{\rm F}T})_{i_2}\cdots ({\bm a}^{\dag}e^{-i{\sf M}_{\rm F}T})_{i_N}|0\>.
\label{eMtrapod}
\eeq
Comparing Eq.~\eqref{utrprod} and \eqref{eMtrapod}, we obtain
\beq
e^{-i{\sf M}_{\rm F}T}=\mathcal{T}e^{-i\int_0^T{\sf M}(t)dt}.
\label{}
\eeq
This equation gives ${\sf M}_{\rm F}$ from ${\sf M}(t)$.

When we consider the limit $L, N\rightarrow\infty$, we immediately know that although the norm of the many-body Hamiltonian diverges, the convergence radius of the Floquet Magnus expansion is finite even in this limit if the norm of the matrix ${\sf M}(t)$ stays finite.

\section{Breaking of the convergence of the Floquet-Magnus expansion}\label{SdfM}
\begin{figure*}
\centering
\subfigure[%
]{
\includegraphics[clip,scale=0.46]{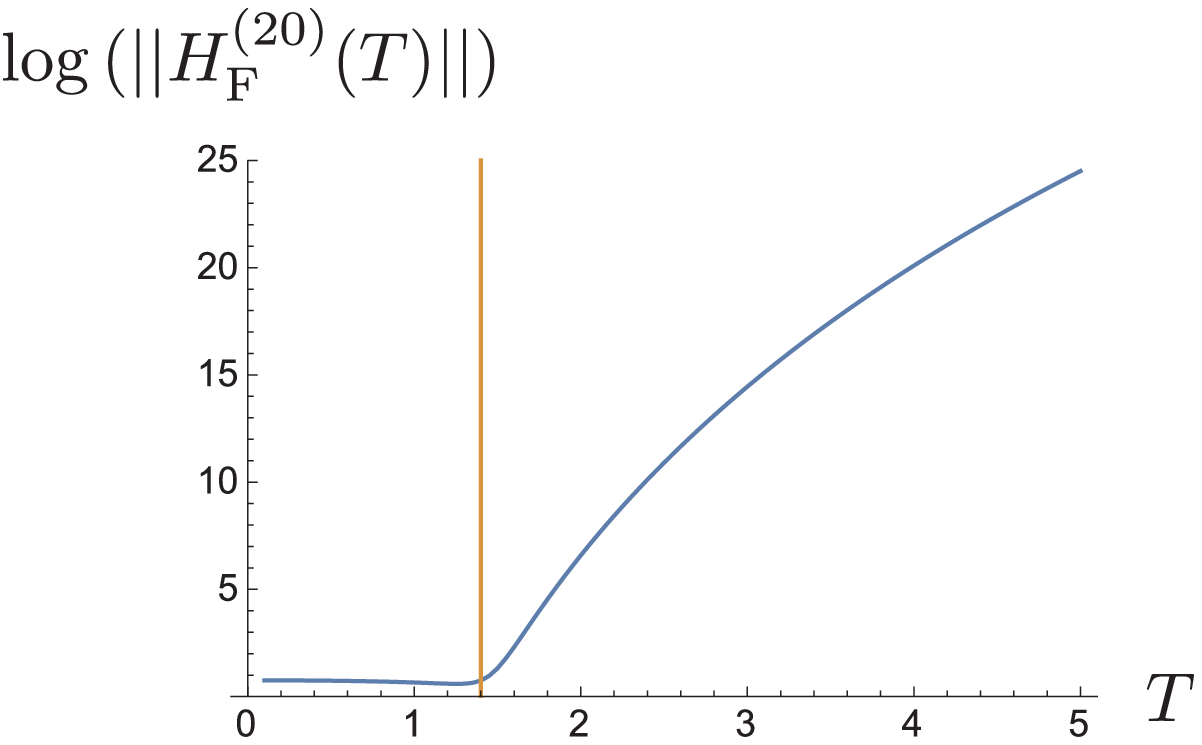}
}
\subfigure[%
]{
\includegraphics[clip,scale=0.46]{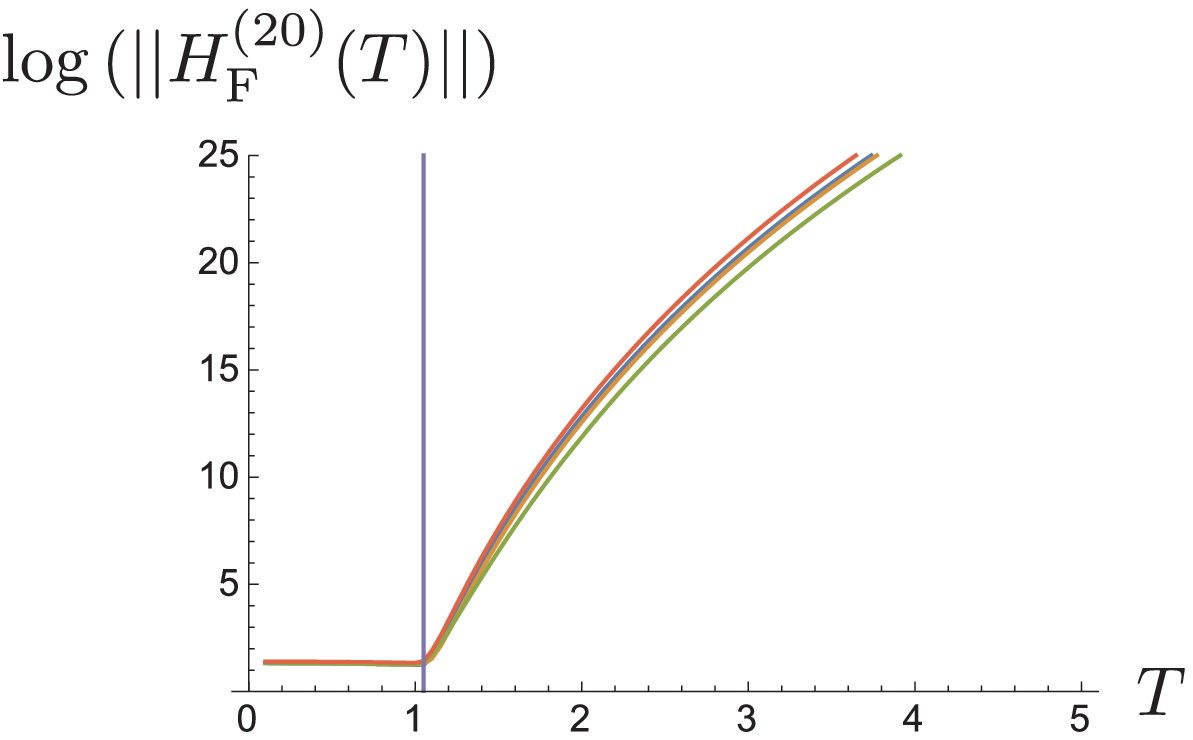}
}
\subfigure[%
]{
\includegraphics[clip,scale=0.46]{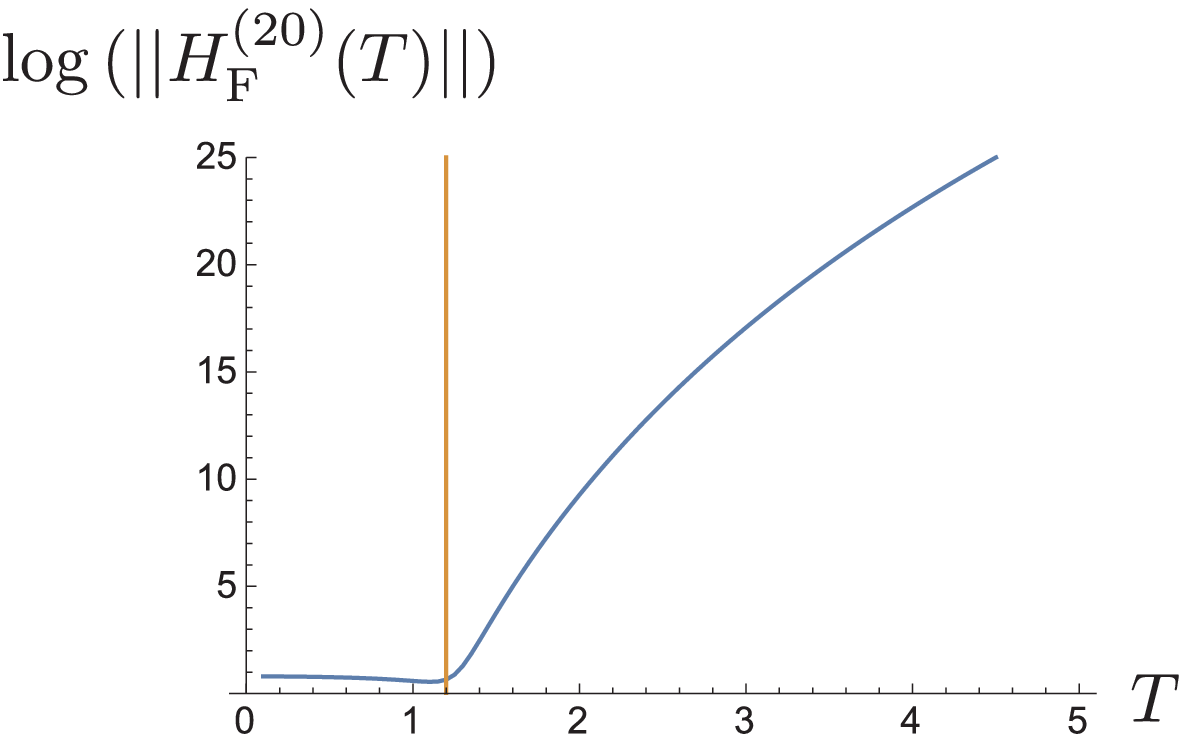}
}
\caption{(color online) Magnitude of $H_{\rm F}^{(20)}(T)$. 
We set $\{h_i\}$ in Eq.~(10) 
 as (a) the quasi-periodic field, (b) the random Gaussian field with four random samples, and (c) the staggered field; see the description below Eq.~(10)
 . 
The system size is $L=500$. The vertical line denotes the driving period where we detected the breaking of the convergence of the expansion. } \label{fig:FMd}
\end{figure*}

In Fig.~\ref{fig:FMd} we show the magnitude of the effective Hamiltonian obtained by the Floquet-Magnus expansion truncated at 20th order, which we denote as $H_{\rm F}^{(20)}(T)=\sum_{n=0}^{20}T^n\Omega_n(T).$ 
We expect this order to be high enough to detect the divergence point of the expansion. 
The vertical line corresponds to the vertical line in Fig.~1
. 
The figures imply that the Floquet-Magnus expansion is divergent when the driving period is larger than the value indicated by the vertical line.

\section{energy density difference and the variance of the effective temperatures for the conserved quantities without scaling}
\begin{figure}
\centering
\subfigure[%
]{
\includegraphics[clip, scale=1]{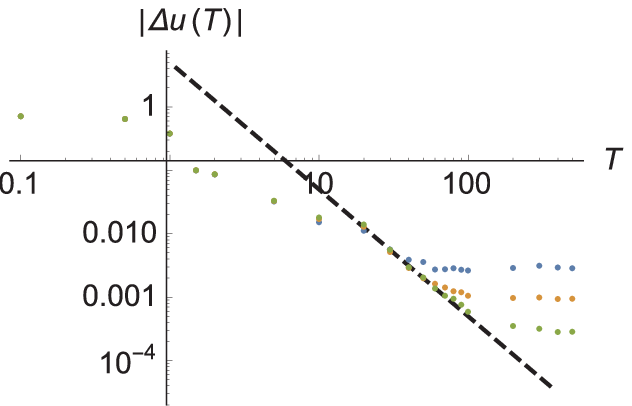}
}
\subfigure[%
]{
\includegraphics[clip,scale=1]{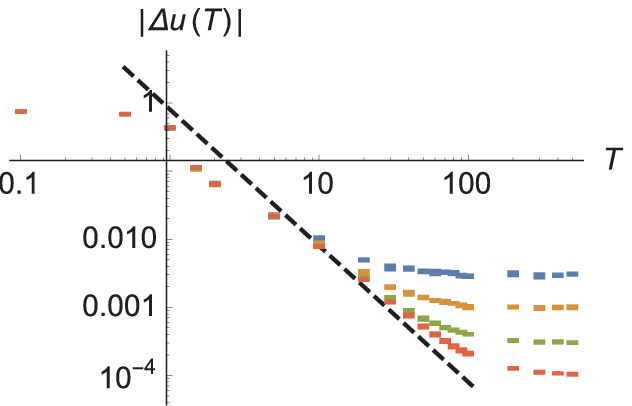}
}
\caption{(color online) Absolute values of the energy-density difference of the steady state. 
We set $\{h_i\}$ in Eq.~(10) 
 as (a) the quasi-periodic fields and (b) the random Gaussian field with four random samples.  
The driving periods for the data points are $T=0.1,\;0.5,\;1,\;1.5,\;2,\;5,\;10,\;20,\;30,\;40,\;50,\;60,\;70,\;80,\;90,\;100,\;200,\;300,\;400,\;500$. The system sizes are the same as in Fig.~2 
 of the main text. 
The larger the system size, the lower lies the data points for large $T$. 
(Different colors also indicate different system sizes.) The broken line indicates the behavior $T^{-2}.$}\label{Fig:ET}
\end{figure}

\begin{figure}
\centering
\subfigure[%
]{
\includegraphics[clip, scale=1]{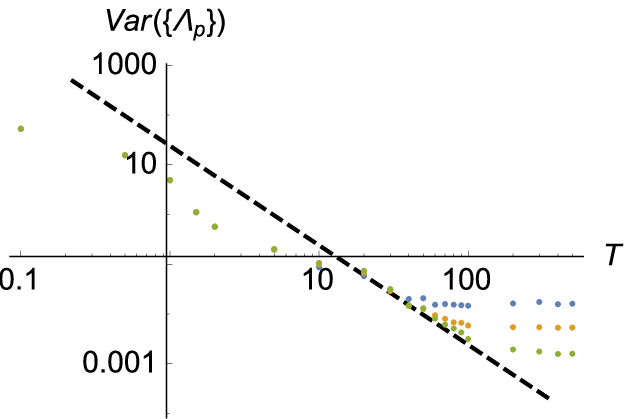}
}
\subfigure[%
]{
\includegraphics[clip,scale=1]{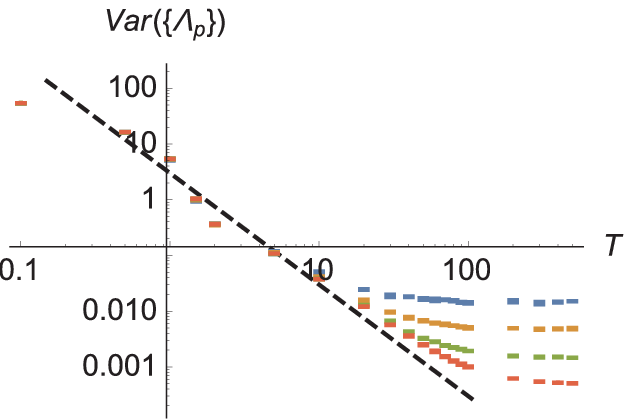}
}
\caption{(color online) Variance of $\{\Lambda_p\}_{p=1}^L$. 
We set $\{h_i\}$ in Eq.~(10) 
 as (a) the quasi-periodic fields and (b) the random Gaussian field with four random samples. 
The driving periods and the system sizes for the data points are the same as in Fig.~\ref{Fig:ET}. The larger the system size, the lower lies the data points for large $T$. 
(Different colors also indicate different system sizes.) The broken line indicates the behavior $T^{-2}.$}\label{Fig:LaT}
\end{figure}

In Figs.~\ref{Fig:ET} and \ref{Fig:LaT}, we show the energy-density difference of the steady state and the variance of $\{\Lambda_p\}_{p=1}^L$ against the driving period, respectively. 
Using these data, we obtained the scaling plots in Fig.~2
.

\end{document}